# Direct evidence of intrinsic Mott state and its layer-parity oscillation in a breathing kagome crystal down to monolayer

Huanyu Liu[1*], Wenhui Li[1,2*], Zishu Zhou[3*], Hongbin Qu[1*], Jiaqi Zhang[1*], Weixiong Hu[1], Chenhaoping Wen[1], Ning Wang[3†], Hao Deng[1,2†], Gang Li[1,2†] and Shichao Yan[1,2†]

[1]*State Key Laboratory of Quantum Functional Materials, School of Physical Science and Technology, ShanghaiTech University, Shanghai 201210, China*
[2]*ShanghaiTech Laboratory for Topological Physics, ShanghaiTech University, Shanghai 201210, China*
[3]*Department of Physics and Centre for Quantum Materials, The Hong Kong University of Science and Technology, Clear Water Bay, Kowloon, Hong Kong, China*

*These authors contributed equally to this work
†Email: phwang@ust.hk; denghao@shanghaitech.edu.cn; ligang@shanghaitech.edu.cn; yanshch@shanghaitech.edu.cn;

**Abstract**
We report direct spectroscopic evidence of correlation-driven Mott states in layered $Nb_3Cl_8$ through combining scanning tunneling microscopy (STM) and dynamical mean-field theory. The Hubbard bands persist down to monolayer, providing the definitive evidence for the Mottness in $Nb_3Cl_8$. While the size of the Mott gap remains almost constant across all layers, a striking layer-parity-dependent oscillation emerges in the local density of states (LDOS) between even ($n$ = 2,4,6) and odd layers ($n$ = 1,3,5), which arises from the dimerization and correlation modulation of the obstructed atomic states, respectively. Our conclusions are supported by a critical technical advance in atomic-scale LDOS mapping for highly insulating systems. This work provides the definitive experimental verification of correlation-driven Mott ground states in $Nb_3Cl_8$ while establishing a general protocol for investigating the interplay of electronic correlation and interlayer coupling in layered insulators by using low-temperature STM technique.

**Main text**

Mott insulator serves as a typical strongly correlated system for exploring exotic many-body phenomena [1,2], including high-temperature superconductivity [3,4], ferromagnetic insulating states [5,6], quantum spin liquids [7-9], and fractionalized excitations [10,11]. In real materials, in addition to intralayer electron-electron interactions, interlayer coupling often plays a significant role and cannot be ignored [12,13]. The competition between these interactions profoundly influences the nature of the ground state, sparking an ongoing controversy in experimental characterization of the insulating ground state in layered correlation systems [12-15]. Specifically, it remains controversial whether these states are Mott insulators induced by intralayer electron correlation or band insulators driven by interlayer coupling.

Recently, $Nb_3X_8$ (X = Cl, Br, I) has emerged as a new class of correlated quasi-two-dimensional (2D) materials, sharing many essential features with transition-metal dichalcogenides, such as 1T-$TaS_2$ [12,16-18]. It was also topologically characterized as an obstructed atomic insulator hosting intrinsic surface states [19], which quickly gained significant attention [20-23]. $Nb_3X_8$ features a clusterization with three Nb atoms forming a trimer in the *ab* plane [Fig. 1(a)]. The shorter intracluster Nb-Nb bonds and longer intercluster Nb-Nb bonds form a breathing kagome lattice [24-27]. The electronic structure of $Nb_3X_8$ is dominated by the Nb 4d orbital [25-29], with one unpaired electron occupying the $2a_1$ orbital on each $Nb_3$ trimer [Fig. 1(b)], nominally resulting in a half-filled metallic band constituted by a molecular orbital that is spatially distributed over the three Nb atoms in each $Nb_3$ trimer [30,31]. As in 1T-$TaS_2$, strong electronic correlations transform this molecular orbital into a cluster Mott insulator [20,23,30,31] [Fig. 1(c)], where electrons are delocalized inside $Nb_3$ trimer but long-range transport is forbidden. This state arises from both local and non-local Coulomb repulsions within the $Nb_3$ trimer [30]. Theoretical calculations suggest that bilayer $Nb_3X_8$ systems exhibit a crossover between Mott and band insulators, depending on the competition between intralayer electron correlation and interlayer dimerization [20]. However, the recent angle-resolved photoemission spectroscopy (ARPES) measurements demonstrate that $Nb_3Br_8$ may be instead a dimerized Mott insulator [23,32] [Fig. 1(d)]. For $Nb_3Cl_8$, ARPES measurements reveal the signature of lower Hubbard band (LHB) in the high-temperature phase where the interlayer coupling is weak [21]. $Nb_3Cl_8$ undergoes a structural phase transition from high-temperature phase (*α*-phase) to low-temperature phase (*β*-phase) at ∼100 K [26,27]. Along the *c*-axis, the unit cell of the *β*-phase consists of six $Nb_3Cl_8$ monolayers stacked with interlayer van der Waals interactions [26]. Adjacent layers adopt two alternating stacking sequences [20,26,33]: A'B stacking configuration for the first and second layers with center-aligned $Nb_3$ trimers, and strong interlayer coupling forms bilayer dimerization [21,26,27]; BB' stacking configuration for the second and third layers with center-staggered $Nb_3$ trimers [Fig. 1(a)].

Given the interlayer dimerization in *β*-phase $Nb_3Cl_8$ [21,26,27], interlayer coupling may lead to a dimerized Mott insulating state. However, the electronic structure of *β*-$Nb_3Cl_8$ cannot be directly obtained by ARPES because of its highly insulating nature [21,34], and $Nb_3Cl_2Br_6$ with higher phase transition temperature has been chosen as a substitute to infer the dimerized Mott insulating state in the *β*-phase $Nb_3Cl_8$. To ambiguously identify the nature of the ground electronic state in low-temperature $Nb_3Cl_8$, it is necessary to systematically investigate the layer-resolved electronic structure, especially for monolayer.

Scanning tunneling microscopy and spectroscopy (STM/STS) techniques can provide layer-dependent information for both occupied and empty electronic states [35-37]. Unfortunately, the highly insulating nature of $Nb_3Cl_8$ also makes the low-temperature STM measurements unfeasible. To overcome this difficulty, we exfoliate bulk $Nb_3Cl_8$ into few-layer flakes and transfer them onto a highly oriented pyrolytic graphite (HOPG) substrate (see Supplemental Material [38], Figs. S1 and S2), which allows electron tunneling by thinning the thickness of $Nb_3Cl_8$ [Fig. 1(e)]. To avoid surface contamination, we encapsulate the few-layer $Nb_3Cl_8$ with graphene, assembling graphene/$Nb_3Cl_8$ heterostructures for subsequent STM measurements.

The typical optical image of the graphene/$Nb_3Cl_8$ heterostructure on the HOPG substrate is depicted in Fig. 1(f). Graphene acts as a transparent tunneling medium originating from its following characteristics: the thinnest thickness, high conductivity, low density of states (DOS), and short decay length of electron wavefunction, which allow the encapsulated $Nb_3Cl_8$ to contribute more electron tunneling probability [36,39]. Figure 1(g) shows the typical atomic-resolution STM topography of the few-layer $Nb_3Cl_8$ sample at 4.6 K, revealing a triangular lattice of uniformly distributed spheres with a periodicity of ~6.5 Å, which matches the lattice of $Nb_3$ trimer [24,30].

To investigate the electronic structure of $Nb_3Cl_8$, we first perform differential conductance (d$I$/d$V$) measurements on graphene/$Nb_3Cl_8$ heterostructure consisting of up to two $Nb_3Cl_8$ layers (bilayer sample) [Figs. 2(a)-2(c)]. The edge of the bilayer sample exposes a portion of monolayer terrace, and the thickness of monolayer $Nb_3Cl_8$ can be inferred to be ~0.6 nm from the step height profile across the bare HOPG substrate and $Nb_3Cl_8$ terrace [insets of Figs. 2(b) and 2(c)], which is consistent with the previous reports [24,40]. As shown in Fig. 2(d), the d$I$/d$V$ spectra of $Nb_3Cl_8$ exhibit remarkable features: both 1L and 2L $Nb_3Cl_8$ exhibit two prominent peaks with similar insulating gap sizes, which is distinct from the spectral features of graphene, i.e., a nearly linear DOS and a small gap of ~ 130 mV arising from the phonon-assisted inelastic electron tunneling [36,39,41]. The two peaks are located above and below the Fermi level, respectively, with a peak-to-peak energy separation of 1.6 eV. These two peaks match the LHB observed by ARPES measurement and the theoretically calculated upper Hubbard band (UHB) for $Nb_3Cl_8$, respectively [21,23,30,31]. Our data indicate that the Mott gap persists down to monolayer limit, where the interlayer interaction is completely removed [Fig. 2(d)]. As shown in the linecut of d$I$/d$V$ spectra on monolayer $Nb_3Cl_8$ [Figs. 2(c) and 2(e)], these two peaks are more or less uniformly spatially distributed, providing compelling evidence that the ground state of monolayer $Nb_3Cl_8$ is a Mott insulating state dominated by strong electron-electron correlations.

The Mott gap feature in the monolayer $Nb_3Cl_8$ covered by graphene can also be reflected in the bias-dependent STM topographies. Outside the Mott gap, the topography predominantly exhibits bright-center $Nb_3$ trimer feature [Fig. 2(f)]. When the bias voltage approaches the Mott gap, the graphene lattice starts to appear, and the STM topography features are a mixture of graphene and $Nb_3$ trimer patterns [Figs. 2(g) and 2(h)]. Whereas within the Mott gap, the topography displays a honeycomb lattice structure of graphene [Fig. 2(i)]. More topographic images of monolayer $Nb_3Cl_8$ at other biases are provided in Supplemental Material [38], Fig. S3. These bias-dependent topographic features once again confirm that the two peaks in the d$I$/d$V$ spectrum originate from the LHB and UHB of monolayer $Nb_3Cl_8$.

In addition to the similar Mott gap size feature for 1L and 2L $Nb_3Cl_8$ [Fig. 2(d)], another noteworthy feature is that the peak intensities of the UHB and LHB of 2L are stronger than those of 1L. This difference reflects variations in the electronic states between different layers, suggesting that the interlayer coupling of $Nb_3Cl_8$ in the low-temperature phase may not be negligible. A reasonable conjecture is that the interlayer dimerization leads to the odd-even layer-dependent electronic states in the low-temperature phase $Nb_3Cl_8$ [21,26,27].

To further verify the conjecture of the odd-even layer-dependent effect, we prepare another graphene/$Nb_3Cl_8$ heterostructure with the $Nb_3Cl_8$ consisting of up to six layers (hexalayer sample) [Fig. 3(a)]. At the edge of the hexalayer sample, the STM topography reveals the exposed $Nb_3Cl_8$ with different layer numbers [Figs. 3(b) and 3(c)], which can be confirmed by the height profile (see Supplemental Material [38], Figs. S4 and S5). The interlayer stacking from 1L to 6L $Nb_3Cl_8$ (Figs. 3(c) and 3(d)) can be identified by replicating and translating the $Nb_3$ trimer lattice from 1L to 2L and 2L to 3L, and so on so force. The observations reveal that the center of $Nb_3$ trimers in 1L and 2L are aligned, while those in 2L and 3L are staggered [Fig. 3(c)]. This kind of stacking order analysis can be applied to the other layers [such as 4L to 6L in Fig. 3(d)]. Our STM data show that the structure of hexalayer $Nb_3Cl_8$ sample adopts the $\beta$-like interlayer stacking arrangement at 4.6 K [21,26,27].

We next perform layer-resolved $dI/dV$ spectrum measurements from 1L to 6L [Figs. 3(a) and 3(b)], and the results are presented in Fig. 3(e) (see also Supplemental Material [38], Figs. S6 and S7). A striking feature is the oscillation of the peak intensities of UHB and LHB for odd and even layers. The peak intensities of different even layers are roughly the same, and significantly stronger than those of the odd layers, which confirms the odd-even layer-dependent effect. Interestingly, the peak-to-peak energy separation between UHB and LHB remains roughly constant at ~1.6 eV for all layers [Fig. 3(g)], indicating that the Mott gap is robust and independent of interlayer coupling. Consequently, these observations imply that the interlayer interactions do not collapse the Mott gap in $Nb_3Cl_8$, which remains a robust Mott insulating state driven primarily by strong electron-electron interactions. Taking together with the odd-even layer-dependent $dI/dV$ spectra, our data show that even-layer $Nb_3Cl_8$ is in the dimerized Mott insulating state. From the $dI/dV$ maps on the 6L $Nb_3Cl_8$ at the peak energies of LHB [−1.1 V, Fig. 3(h)] and UHB [0.5 V, Fig. 3(i)], we can see that they exhibit similar spatial distributions [Fig. 3(j)], which is consistent with the fact that they originate from the same electronic band (more $dI/dV$ maps are shown in Supplemental Material [38], Fig. S8). We note that a higher-energy electronic peak appears above UHB for all layers [Fig. 3(e)]. Its existence in the monolayer suggests that the higher-energy peak is not due to the splitting of UHB induced by interlayer coupling. However, the intensity of the higher-energy peak also exhibits layer-number dependence, indicating this electronic state may also be affected by the interlayer dimerization effect.

As the layer number of $Nb_3Cl_8$ changes from 1L to 6L, the peak energy positions of UHB and LHB slightly shift toward lower energies [Fig. 3(f)]. Additionally, we find that the peak energies of UHB and LHB in the 2L for the bilayer sample and the 6L for the hexalayer sample remain almost the same [Fig. 3(e)]. This could be because for different samples, the topmost $Nb_3Cl_8$ layer always has the largest terrace which results in similar interaction with the encapsulated graphene. For other layers in the hexalayer sample, their exposed areas are relatively small and the coupling with the encapsulated graphene is different, which makes the energy positions of the UHB and LHB vary for different layers [36,37,39,42] [Fig. 3(f)]. We also notice that the energy positions of UHB and LHB in the 1L of the bilayer and hexalayer samples are also different [Figs. 2(d) and 3(e)]. This may be due to the different interfaces between 1L $Nb_3Cl_8$ and HOPG substrate for different samples (see Supplemental Material [38], Fig. S9). Nevertheless, for all the $Nb_3Cl_8$ layers of both samples, the peak-to-peak energy separation between UHB and LHB remains more or less constant at ~1.6 eV.

To elucidate the evolution of the Mott gap with the layer number and the layer-parity dependent electronic states, we further carry out density functional theory (DFT) calculations for the electronic band structure of 1L-6L $Nb_3Cl_8$ constructed in A'-B-B'-C-C'-A order [Fig. 1(a)]. As evident in Figs. 4(a)-4(f), without accounting for electron-electron interactions, a single band crosses the Fermi level in the odd layers consistent with their obstructed atomic insulating nature [19], whereas a charge gap opens at the Fermi level in even layers, indicating the significance of the interlayer coupling and the emergence of dimerization. Due to the breaking of either A'B, B'C, or C'A dimerization, the electronic states of odd-layer samples lose their partner in one layer, therefore, creating the obstructed surface states as a flat band crossing the Fermi level. When electron-electron interactions are included, dynamical mean-field theory (DMFT) calculations for odd-layer samples reveal that the half-filled band splits into UHB and LHB [Fig. 4(g)], thereby creating the Mott gap. For the even-layer samples, electronic correlations change the bonding/antibonding gap to a correlated Mott gap [32] [Fig. 4(h)]. The size of the Mott gap remains nearly constant regardless of the sample thickness, but the local density of states (LDOS) in the bilayer accounts for two bands below the Fermi level, making it twice as strong as that in the monolayer.

To better compare with the d$I$/d$V$ spectra, we adjusted the DMFT chemical potential and the Coulomb repulsion $U = 1.5$ eV to match the location of UHB and LHB with the experiment [right panels in Figs. 4(g) and 4(h)]. The Mott gap is found to be much larger than the band gap extracted from DFT calculation, further corroborating that the ground state of $Nb_3Cl_8$ is a dimerized Mott phase. The fine splitting of the LHB and UHB in DMFT calculations shown in the colored LDOS in Fig. 4(h) is due to the transition from band insulator to Mott insulator induced by the strong electronic correlations [20,21,43], and it is not clearly resolved in the d$I$/d$V$ spectrum. This is likely due to the finite momentum dispersion of the Hubbard bands and the lack of momentum resolution in the d$I$/d$V$ measurements, resulting in a single electronic peak with higher intensity [Fig. 4(h)]. DMFT calculations for the LDOS in the trilayer and quadlayer $Nb_3Cl_8$ are provided in Supplemental Material [38], Fig. S10. Since the majority of signals in STM measurements originate from the topmost surface, for thicker samples, the d$I$/d$V$ spectra mainly reflect the oscillation of the top monolayer and bilayer in odd- and even-layer samples, respectively. This leads to the oscillation of the d$I$/d$V$ signal while the gap size remains fixed. Our theoretical finding is highly consistent with the experiment.

In summary, by developing an advanced experimental protocol, we manage to measure the frustrated insulating state of $Nb_3Cl_8$ with low-temperature STM, which provides direct evidence for the controversial nature of the ground states. We convincingly show that the ground state of monolayer $Nb_3Cl_8$ is a Mott insulator driven by strong electronic correlations. Interlayer coupling mediates the Mott states in multilayer samples and creates a layer-parity oscillation in the LDOS. This layer-parity oscillation behavior is expected to be observed in other cluster Mott insulators with strong interlayer coupling, such as 1T-TaS$_2$. These findings redefine the role of layer engineering in 2D correlated systems. The observed interplay between on-site Coulomb interaction and interlayer dimerization suggests that Nb$_3$X$_8$ family may host exotic excitations at parity interfaces, offering a tunable route to correlation-mediated topological states. Our work also paves the way for extending STM-based diagnostics to broader classes of insulating quantum matter. In principle, this method can also be extended to study flat-band induced correlation effects and interlayer coupling in classic kagome systems.

**Acknowledgments**

S.Y., G.L. and H.D. acknowledge the financial support from the National Key Research and Development Program of China (Grant No. 2022YFA1402703) and the start-up funding from ShanghaiTech University. S.Y. and N.W. acknowledge the financial support from the National Key Research and Development Program of China (Grant No. 2020YFA0309602). H.D. acknowledges the financial support from Science and Technology Commission of Shanghai Municipality (Grant No. 21PJ1410000). W.L. acknowledges the financial support from the National Natural Science Foundation of China (Grant No. 12404222), the Postdoctoral Fellowship Program of CPSF (Grant No. GZC20231670) and the China Postdoctoral Science Foundation (Grant No. 2024M762085).

**References**

[1] J. Hubbard, Proc. R. Soc. A **276**, 238 (1963).
[2] N. F. Mott, Rev. Mod. Phys. **40**, 677 (1968).
[3] F. C. Zhang and T. M. Rice, Phys. Rev. B **37**, 3759 (1988).
[4] P. A. Lee, N. Nagaosa, and X.-G. Wen, Rev. Mod. Phys. **78**, 17 (2006).
[5] J. Cai *et al.*, Nature **622**, 63 (2023).
[6] P. Mai, B. E. Feldman, and P. W. Phillips, Phys. Rev. Res. **5**, 013162 (2023).
[7] K. W. Plumb, J. P. Clancy, L. J. Sandilands, V. V. Shankar, Y. F. Hu, K. S. Burch, H.-Y. Kee, and Y.-J. Kim, Phys. Rev. B **90**, 041112 (2014).
[8] A. Koitzsch *et al.*, Phys. Rev. Lett. **117**, 126403 (2016).
[9] A. Banerjee *et al.*, Science **356**, 1055 (2017).
[10] G. Chen, H.-Y. Kee, and Y. B. Kim, Phys. Rev. Lett. **113**, 197202 (2014).
[11] U. F. Seifert, X.-Y. Dong, S. Chulliparambil, M. Vojta, H.-H. Tu, and L. Janssen, Phys. Rev. Lett. **125**, 257202 (2020).
[12] Y. Wang, W. L. Yao, Z. M. Xin, T. T. Han, Z. G. Wang, L. Chen, C. Cai, Y. Li, and Y. Zhang, Nat. Commun. **11**, 4215 (2020).
[13] C. Wen *et al.*, Phys. Rev. Lett. **126**, 256402 (2021).
[14] Y. Chen *et al.*, Nat. Phys. **16**, 218 (2020).
[15] J. Lee, K.-H. Jin, and H. W. Yeom, Phys. Rev. Lett. **126**, 196405 (2021).
[16] J.-J. Kim, W. Yamaguchi, T. Hasegawa, and K. Kitazawa, Phys. Rev. Lett. **73**,

2103 (1994).

[17] J.-J. Kim, I. Ekvall, and H. Olin, Phys. Rev. B **54**, 2244 (1996).

[18] C. J. Butler, M. Yoshida, T. Hanaguri, and Y. Iwasa, Nat. Commun. **11**, 2477 (2020).

[19] Y. Xu, L. Elcoro, Z.-D. Song, M. Vergniory, C. Felser, S. S. Parkin, N. Regnault, J. L. Mañes, and B. A. Bernevig, Phys. Rev. B **109**, 165139 (2024).

[20] Y. Zhang, Y. Gu, H. Weng, K. Jiang, and J. Hu, Phys. Rev. B **107**, 035126 (2023).

[21] S. Gao *et al.*, Phys. Rev. X **13**, 041049 (2023).

[22] S. Regmi *et al.*, Phys. Rev. B **108**, L121404 (2023).

[23] M. Date *et al.*, Nat. Commun. **16**, 4037 (2025).

[24] S. N. Magonov, P. Zönnchen, H. Rotter, H. J. Cantow, G. Thiele, J. Ren, and M. H. Whangbo, J. Am. Chem. Soc. **115**, 2495 (1993).

[25] M. Ströbele, J. Glaser, A. Lachgar, and H. Meyer, Z. Anorg. Allg. Chem. **627**, 2002 (2001).

[26] Y. Haraguchi, C. Michioka, M. Ishikawa, Y. Nakano, H. Yamochi, H. Ueda, and K. Yoshimura, Inorg. Chem. **56**, 3483 (2017).

[27] J. P. Sheckelton, K. W. Plumb, B. A. Trump, C. L. Broholm, and T. M. McQueen, Inorg. Chem. Front. **4**, 481 (2017).

[28] J. R. Kennedy, P. Adler, R. Dronskowski, and A. Simon, Inorg. Chem. **35**, 2276 (1996).

[29] G. Chen, H.-Y. Kee, and Y. B. Kim, Phys. Rev. B **93**, 245134 (2016).

[30] J. Hu, X. Zhang, C. Hu, J. Sun, X. Wang, H.-Q. Lin, and G. Li, Commun. Phys. **6**, 172 (2023).

[31] S. Grytsiuk, M. I. Katsnelson, E. G. v. Loon, and M. Rösner, npj Quantum Materials **9**, 8 (2024).

[32] J. Aretz *et al.*, Preprint at https://arxiv.org/abs/2501.10320 (2025).

[33] C. M. Pasco, I. El Baggari, E. Bianco, L. F. Kourkoutis, and T. M. McQueen, ACS Nano **13**, 9457 (2019).

[34] J. Yoon, E. Lesne, K. Sklarek, J. Sheckelton, C. Pasco, S. S. Parkin, T. M. McQueen, and M. N. Ali, J. Phys. Condens. Matter **32**, 304004 (2020).

[35] C. Ye, P. Cai, R. Yu, X. D. Zhou, W. Ruan, Q. Q. Liu, C. Q. Jin, and Y. Y. Wang, Nat. Commun. **4**, 1365 (2013).

[36] Z. Qiu *et al.*, Nat. Mater. **23**, 1055 (2024).

[37] X. Zheng, Z.-X. Liu, C. Zhang, H. Zhou, C. Yang, Y. Shi, K. Tanigaki, and R.-R. Du, Nat. Commun. **15**, 7658 (2024).

[38] See Supplemental Material at http://link.aps.org/supplemental for detailed experimental methods and DFT calculations, as well as relevant experimental data and analysis, which includes Refs. [30,36,39].

[39] Z. Qiu *et al.*, Nat. Commun. **12**, 70 (2021).

[40] B. Lee *et al.*, Nanoscale **16**, 20312 (2024).

[41] Y. B. Zhang, V. W. Brar, F. Wang, C. Girit, Y. Yayon, M. Panlasigui, A. Zettl, and M. F. Crommie, Nat. Phys. **4**, 627 (2008).

[42] N. Tilak *et al.*, Nat. Commun. **15**, 8056 (2024).

[43] C. Hu, H. Qu, X. Zhang, X.-Q. Wang, H.-Q. Lin, and G. Li, Phys. Rev. B **110**,

**Figures**

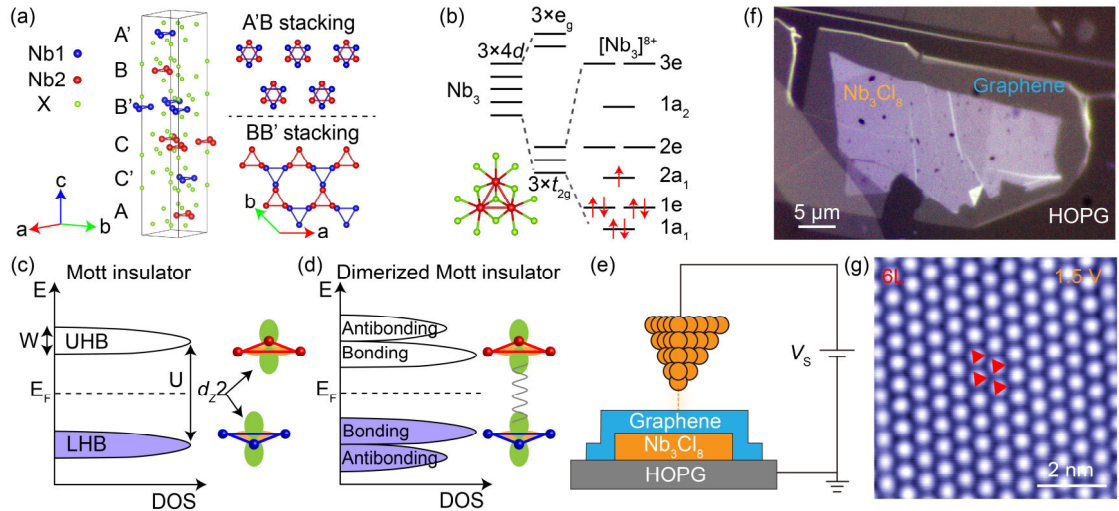

FIG. 1 (a) Left panel: the schematic crystal structure of $Nb_3X_8$ at low temperature (side view). Right panel: adjacent layers of $Nb_3$ trimer adopt an alternating stacking configuration with center-aligned and staggered arrangements (top view). (b) Schematic view of the molecular orbitals of the $Nb_3X_{13}$ cluster unit containing a $Nb_3$ trimer with seven valence electrons. (c),(d) Sketch of Mott insulator and dimerized Mott insulator, respectively. (e) Schematic illustration of STM measurement setup. (f) Optical image of graphene/$Nb_3Cl_8$ heterostructures on HOPG substrate. (g) Constant current STM topography of 6L $Nb_3Cl_8$ taken at sample bias voltage $V_s$ = 1.5 V and tunneling current $I_t$ = 100 pA. Red triangles indicate $Nb_3$ trimers.

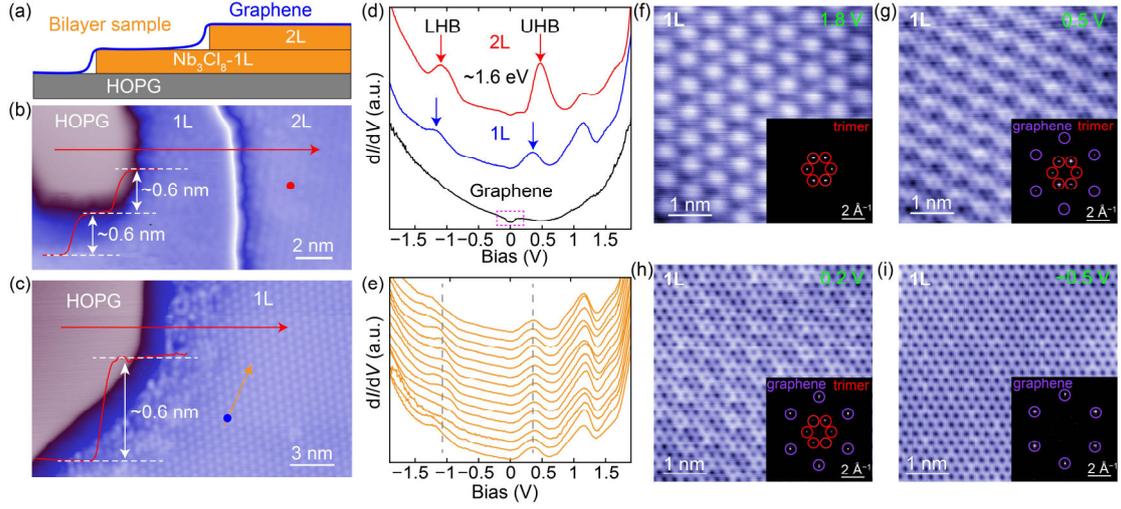

FIG. 2 (a) Schematic illustration of bilayer sample. (b),(c) STM topography of bilayer and monolayer $Nb_3Cl_8$. (b) $V_s$ = 1.9 V, $I_t$ = 50 pA; (c) $V_s$ = 2 V, $I_t$ = 50 pA. Step height profiles are taken along the red arrows. (d) d$I$/d$V$ spectra of bilayer $Nb_3Cl_8$ and graphene. The d$I$/d$V$ spectra of $Nb_3Cl_8$ were measured at the marked spots in (b) and (c). The arrows indicate the peaks of UHB and LHB. The purple dashed rectangle marks the inelastic tunneling gap feature from graphene. (e) Linecut of d$I$/d$V$ spectra on monolayer $Nb_3Cl_8$ along the yellow arrow in (c). The grey dashed lines mark the peaks of UHB and LHB. The spectra in (d) and (e) are vertically offset for clarity. (f-i) Bias-dependent of STM topographies on monolayer $Nb_3Cl_8$. The insets indicate their corresponding Fourier transform images of STM topographies.

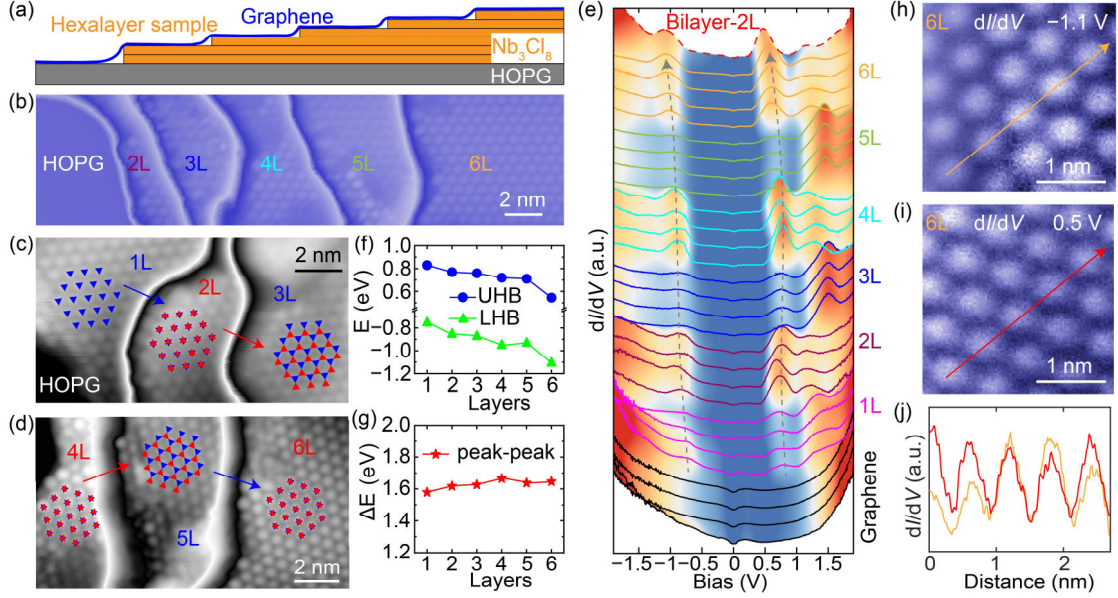

FIG. 3 (a) Schematic illustration of hexalayer sample. (b) The STM topography of hexalayer $Nb_3Cl_8$ sample ($V_s$ = 1 V, $I_t$ = 10 pA). (c) The STM topography of 1L-3L $Nb_3Cl_8$, where red and blue triangles denote $Nb_3$ trimers in even and odd layers, respectively. The $Nb_3$ trimer pattern is replicated and translated from 1L to 2L and 2L to 3L, respectively, to determine interlayer stacking configuration ($V_s$ = 2 V, $I_t$ = 30 pA). (d) The STM topography of 4L-6L $Nb_3Cl_8$. The $Nb_3$ trimer pattern is replicated and translated from 4L to 5L and 5L to 6L, respectively ($V_s$ = 2 V, $I_t$ = 20 pA). (e) The d$I$/d$V$ spectra for different $Nb_3Cl_8$ layers in the hexalayer sample and graphene. The d$I$/d$V$ spectra are taken on different layers in (b) and (c). The dotted arrow indicates the trend in peak energies of UHB and LHB. The red dotted spectrum is the d$I$/d$V$ spectrum on the 2L $Nb_3Cl_8$ of the bilayer sample. The spectra are vertically offset for clarity. (f) Peak energy positions of UHB and LHB for each layer in the hexalayer $Nb_3Cl_8$. (g) Layer-dependent peak-to-peak energy between UHB and LHB. (h),(i) The d$I$/d$V$ map of the 6L taken at LHB and UHB, respectively. (j) The spatially dependent d$I$/d$V$ intensity plotted along the arrows in (h) and (i).

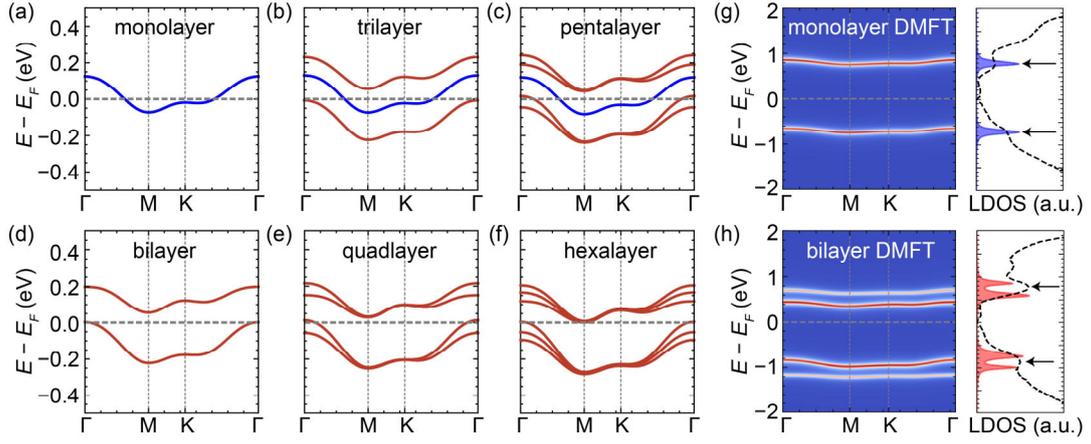

FIG. 4 (a-f) The electronic band structure of 1L-6L $Nb_3Cl_8$ using DFT. The blue curve corresponds to the band from the monolayer or the undimerized single layer. The red bands are the bonding/antibonding bands of the dimerized bilayers. (g),(h) The electronic band structure and corresponding LDOS for monolayer and bilayer $Nb_3Cl_8$ using DMFT, respectively. The DMFT LDOS is shown as filled color, which agrees well with the d$I$/d$V$ curves shown as black dashed lines.